\documentclass[mathleft]{an}
\usepackage{graphicx}
\begin{document}

\Pagespan{789}{}
\Yearpublication{2006}%
\Yearsubmission{2005}%
\Month{11}%
\Volume{999}%
\Issue{88}%

\title{A web-tool for population synthesis of near-by cooling neutron stars: An on-line test for cooling curves}

\author{P.A. Boldin\inst{1}
\and  S.B. Popov\inst{2}\fnmsep\thanks{Corresponding author:
  \email{polar@sai.msu.ru}\newline}
\and N. Tetzlaff\inst{3}
}
\titlerunning{Web-tool for population synthesis of cooling neutron stars}
\authorrunning{P.A. Boldin, S.B. Popov \& N. Tetzlaff}
\institute{
National Research Nuclear University ``Moscow Enginering Physics Institute'', Kashirskoe shosse 31, Moscow 115409, Russia
\and 
Sternberg Astronomical Institute, Universitetski pr. 13,
Moscow 119991, Russia
\and 
Astrophysikalisches Institut und Universit\"ats-Sternwarte Jena, Schillerg\"asschen 2-3, 07745 Jena, Germany}

\received{}
\accepted{}
\publonline{}


\abstract
{We present a new web-tool -- \textsc{Net-PSICoNS} -- for population
synthesis of isolated near-by cooling neutron stars (NSs).
The main aim is to provide an easy test of models of the NS thermal evolution which can be used by groups studying this subject. 
A user can upload cooling curves for a set of
 masses, modify the mass spectrum if necessary, change radii to fit the EoS used for cooling curve calculations, 
and then a population synthesis of close-by isolated cooling NSs is performed. The output includes the Log N
-- Log S distribution confronted with the ROSAT observations
and several other sets of data. In this paper, we summarize the idea of the test proposed by Popov et al. (2006), and
present a user's manual for the web-tool.}

\maketitle

\section{Introduction}

Neutron stars (NSs)
\footnote{Here we refer to NSs as all types of compact objects like hybrid stars, hyperon stars, quark
stars, etc.}
are interesting not only as extreme astrophysical sources 
but also as physical labs with extreme conditions unavailable on Earth: 
superstrong gravity, superstrong magnetic fields, superfluidity and
superconductivity,  etc. (see the book by Haensel et al. 2007, presently, the most complete volume 
on NS physics and astrophysics). 
In this list, a leading role belongs to supernuclear densities in NS interiors 
(about this subject see a detailed review in Schmitt 2010, and a more popular introduction can be found in Pizzochero 2010).  
The equation of state (EoS) for the matter inside NSs
is not known at present, despite many theoretical efforts have been made. The problem is partly related
to the very complicated nature of processes at such high densities. Partly, we are dealing with lack
of observational data. 

One of the most direct methods to confront theoretical predictions on the EoS with data is to use the mass-radius diagram
(see, for example, a review in Prakash et al. 2001).
Potentially, one very precise measurement of the radius and mass for the {\it same} object would solve
the problem. But at the moment no such measurements exist. Because of that astrophysicists are looking for other ways
to compare theoretical predictions with observational data on NSs. One  promising approach is related
to studies of the thermal evolution.\footnote{In addition, cooling properties of NSs are very interesting in respect with 
neutrino processes at high density.} In this case, theoretical cooling curves, which are dependent upon the internal properties
of compact objects,  must be somehow confronted with 
observational data on cooling NSs. Also cooling curves can depend on
different mechanisms of re-heating of a NS, for example, magnetic field
decay (see, for example, Pons, Geppert 2007). This dependence can be studied
by comparison with the same data, too.

The most simple approach  is just a comparison of cooling curves (temperature vs. age: $T-t$)
with data on NSs with known ages and surface temperatures. The main problems are the following. At first,
just for
several objects
there are measurements of both parameters ($T, t$) with high precision (in
the case of significant magnetic field decay characteristic ages of NSs can
be very different from their real ages). Then, in many cases, the obtained values are strongly model
dependent (for example, due to a NS atmosphere). Finally, selection effects can be important, so cooling can be studied
in a very limited range of ages and temperatures. To breakthrough, it is necessary to use additional tests, which would use
other data sets and different populations of objects. 

One additional test was proposed by Popov et al. (2006). The idea is to use a population synthesis of near-by isolated cooling NSs (ICoNSs)
 to confront
calculations of the thermal evolution with observations. In the next section we describe this approach. 
It is important, in our opinion, to provide for the community an opportunity to run the test on the Web, just uploading cooling curves
and specifying additional necessary parameters.
Web-tools for different astrophysical applications are numerous and useful. In the field of NSs it is necessary to mention
the tool to model the radio pulsar survey (Ridley \& Lorimer 2010). In this paper, we present a tool for cooling NSs.
In section 3 we present a user's manual for the web-tool based on the population synthesis of close-by  ICoNSs.

\section{A Log N -- Log S test for cooling curves} 

 A new test for cooling curves of compact objects was proposed by Popov et al. (2006a). 
It is based on the population synthesis (Popov \& Prokhorov 2007) of close-by ICoNSs developed since 2003
(Popov et al. 2003, 2005, 2006b, 2010; Posselt et al. 2008), and we refer to these papers for details.

\begin{figure}
\includegraphics[width=80mm]{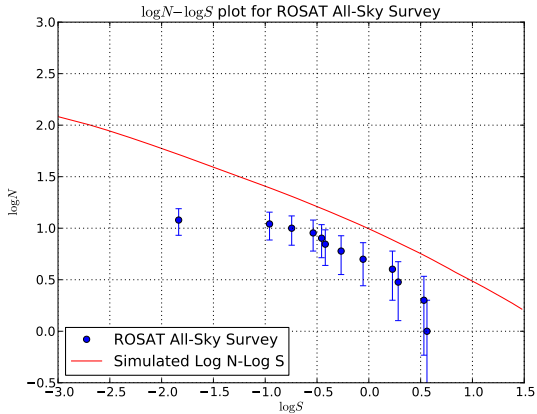}
\caption{Log N -- Log S distribution, calculated using the cooling curves from Blaschke et al. (2004).
This example is included in the demo-version of the code.}
\label{label1}
\end{figure}

 Cooling curves are among the main ingredients of the population synthesis of ICoNSs. Other model assumptions (initial
spatial distribution, velocity distribution, ISM properties, NS mass spectrum etc.) typically are better constrained than 
the thermal evolution. Then, comparison between the calculated and observed populations provides a possibility to test
the worst defined ingredient of the model -- cooling curves. 

 The observed population of ICoNSs is not large. It includes three normal radio pulsars with detected thermal emission
(PSR B1055-52, PSR B0833-45, PSR B0656+14), Geminga (PSR B0633+17) with the so-called ``second Geminga'' (3EG J1835+5918), and
seven radio quiet NSs called the Magnificent Seven (M7, see a review in Haberl 2007). Calvera can be a new member of this
group if indeed its distance is below 1 kpc, as proposed by Zane et al. (2010). On the other hand, PSR B1055-52 
can be slightly further away than the others, and should be excluded from the population of  close-by ICoNSs.  This is a relatively
uniform population, detected by one device (ROSAT) in an all-sky survey.  So, it allows to make a comparison with calculations if a 
good approach is chosen. In our opinion, the best one is to use the Log~N~--~Log~S distribution. 

The effectiveness of the test was demonstrated in Popov et al. (2006a, 2006b), where the code has been run for many models
of the thermal evolution for different EoS. The test has several advantages:
it can be applied to sources with unknown ages
and temperatures, it is sensitive (in the case of the population in the solar vicinity) to relatively large NS ages.

It was shown that even for models which successfully pass other tests (including the most direct
temperature-age test) still can fail in explaining the Log~N~--~Log~S. The latter cannot supersede the $T-t$ test (note, that
there are also several weaker additional tests and considerations: ``mass constraint'', ``brightness constraint''), 
but clearly can work as an important additional test.

\subsection{The population synthesis code: input and output}

 The Web-tool \textsc{Net-PSICoNS} is based on the latest version of the population synthesis code (Popov et al. 2010). 
Here we briefly describe it. More details can be found, for example, in Posselt et al.
(2008).

We calculate the evolution of young (typically $\sim$1 Myr) close-by (distance from the Sun at birth $<$3 kpc) NSs.
The main ingredients of the model are:

\begin{itemize}
\item Initial spatial distribution of NSs;
\item Initial (kick) velocity distribution;
\item Mass spectrum;
\item Cooling curves;
\item The Galactic potential;
\item ISM distribution;
\item Absorbing properties of the ISM;
\item Detector properties.
\end{itemize}

We assume that the initial spatial distribution of NSs follows that of massive stars. 
In the 3 kpc solar neighbourhood it is dominated by the Gould Belt (see P\"oppel 1997) and farther OB-associations. 
According to our standard assumption, 270 NSs are born inside 3 kpc in one million year.
If a user prefers a different normalization, then the output Log N -- Log S
can be easily scaled, as the birth rate works just as a multiplicative
factor.

\begin{figure}
\includegraphics[width=76mm]{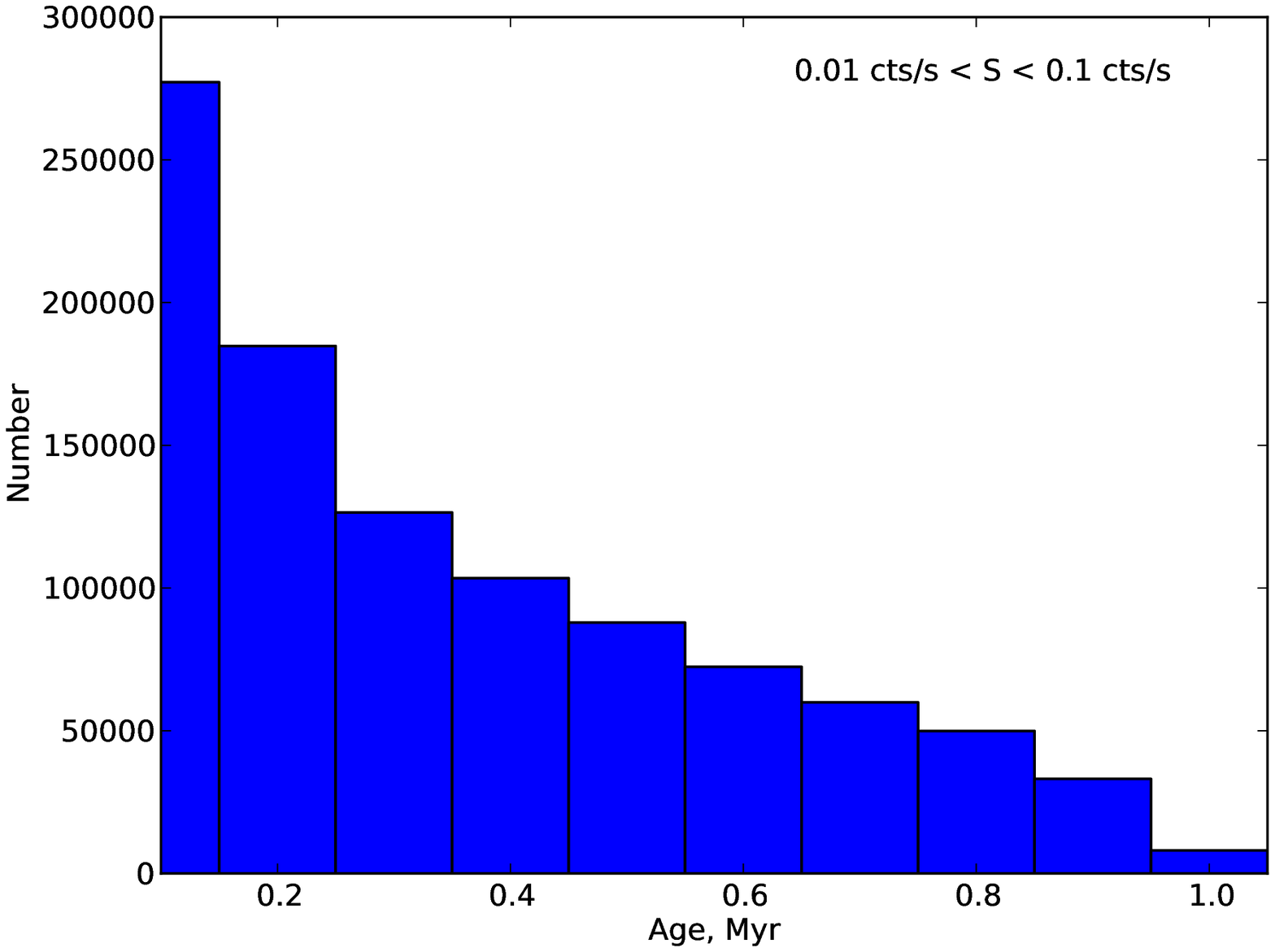}
\includegraphics[width=76mm]{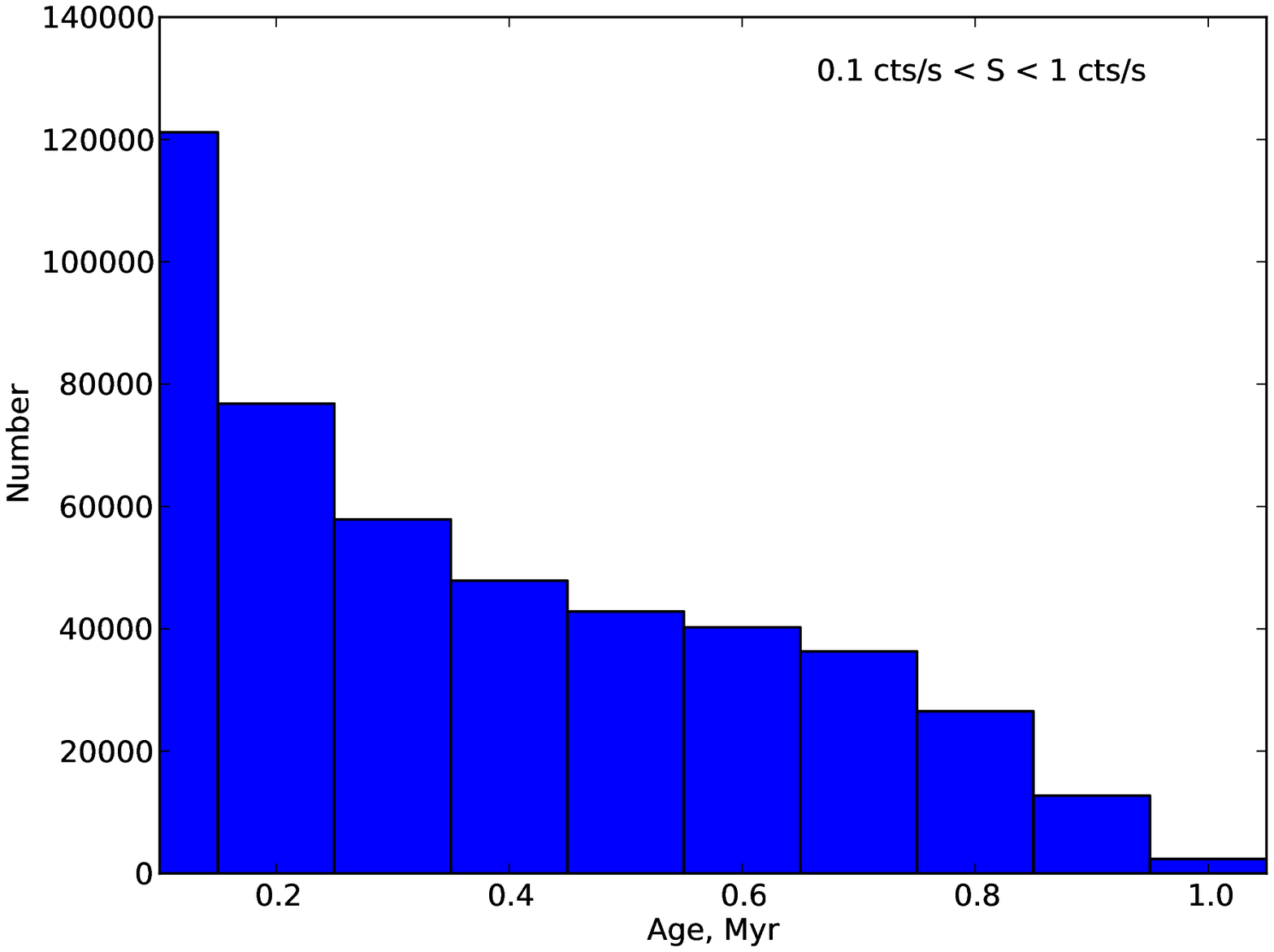}
\includegraphics[width=76mm]{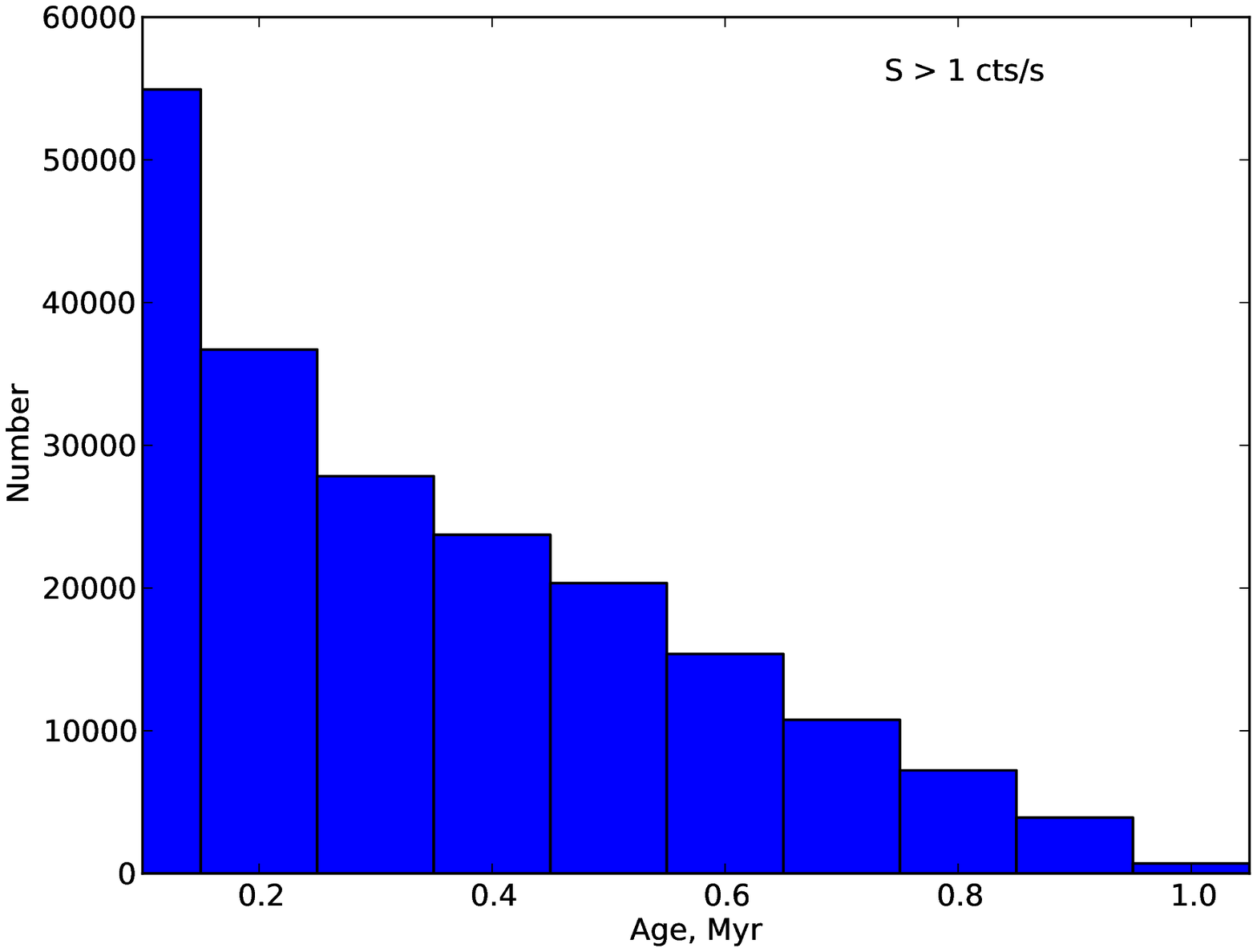}
\caption{Age distribution for three ranges of count rates for the same run as in Figure 1. 
Top panel: $0.01<S<0.1$ cts~s$^{-1}$.
Middle panel: $0.1<S<1$ cts~s$^{-1}$. Bottom panel: $1<S<10$
cts~s$^{-1}$. This example is included in the demo-version of the code.}
\label{label2}
\end{figure}

\begin{figure}
\includegraphics[width=76mm]{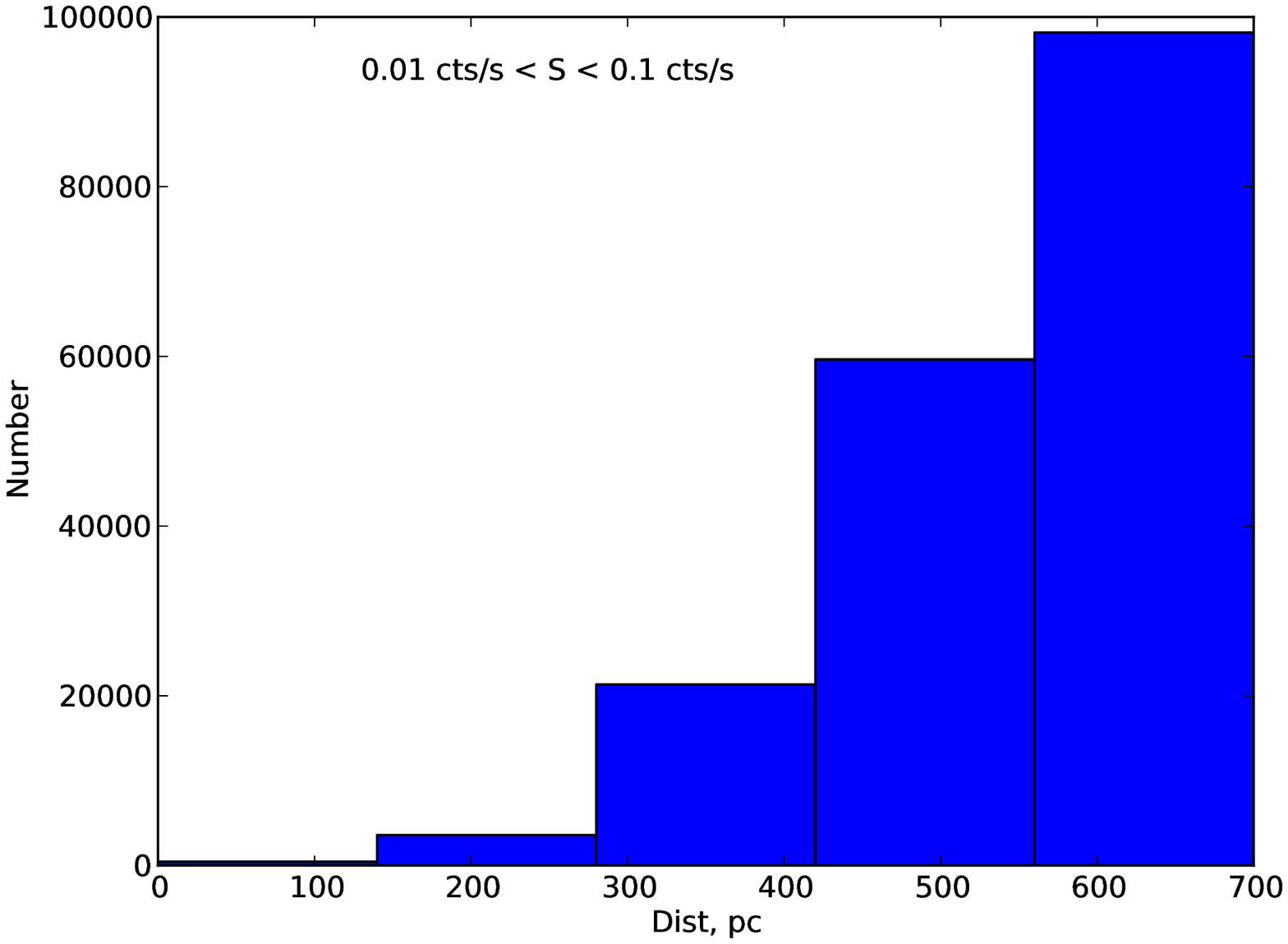}
\includegraphics[width=76mm]{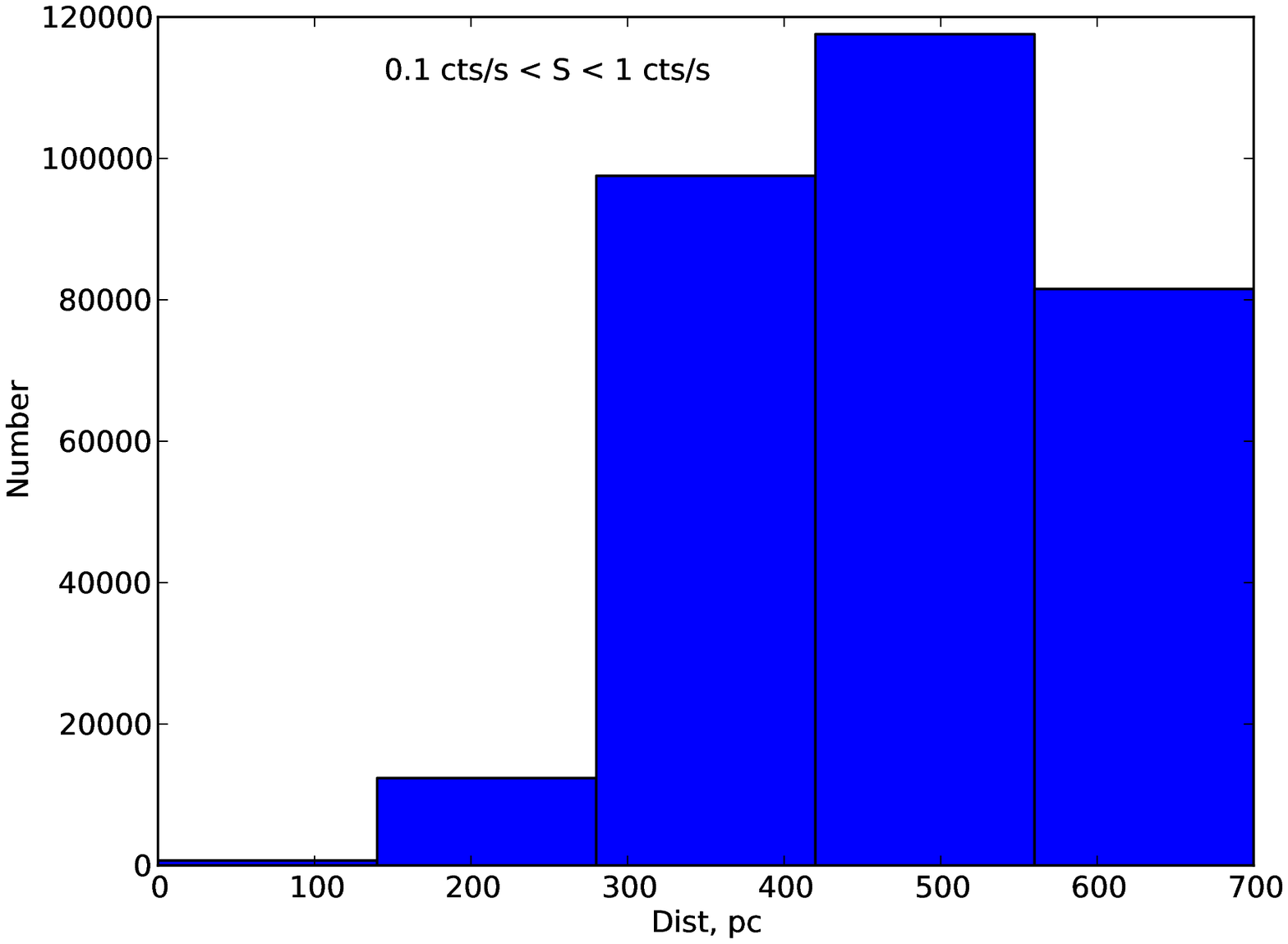}
\includegraphics[width=76mm]{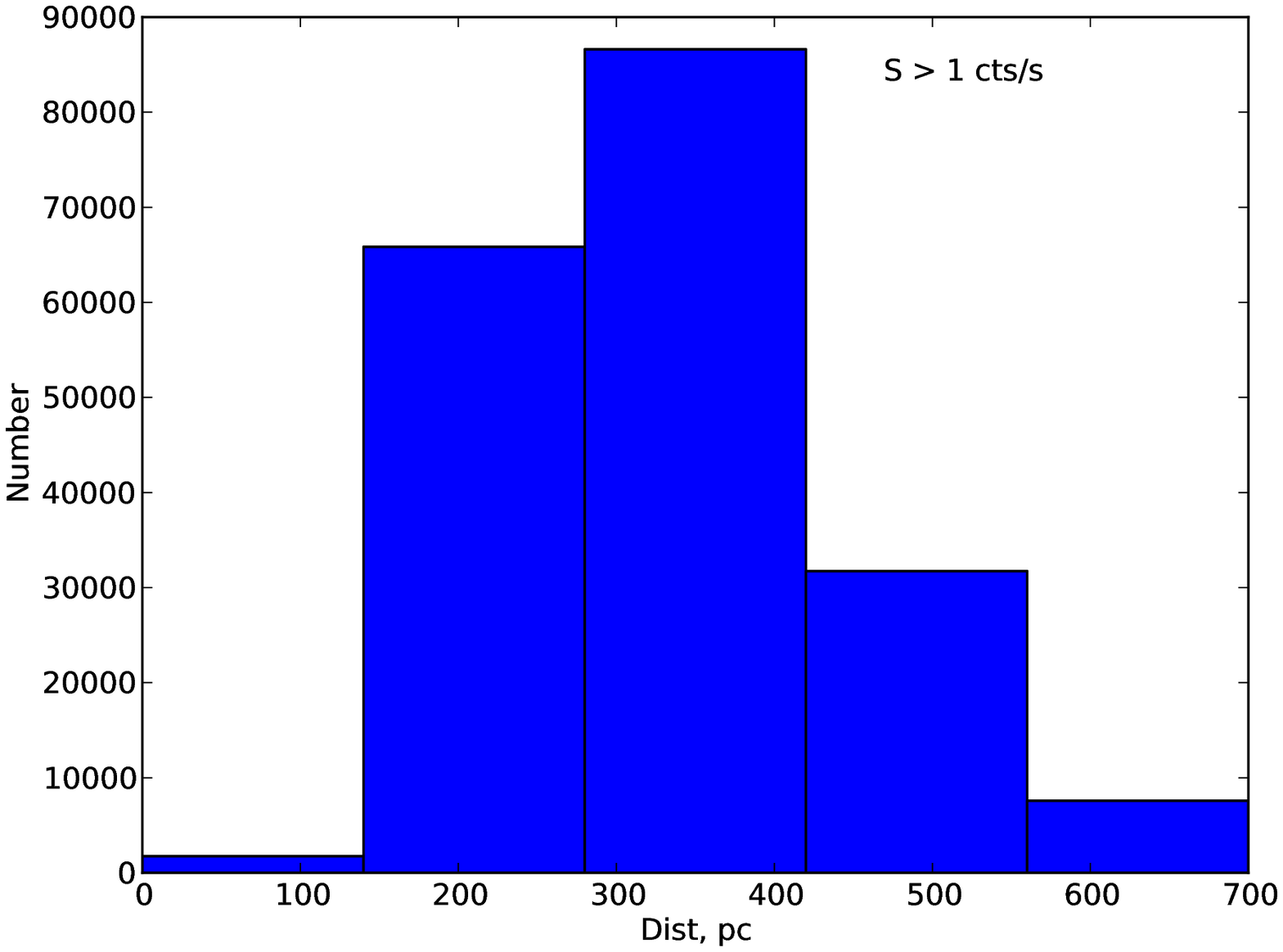}
\caption{Distance distribution for three ranges of count rates for the same run as in Figure 1. Top panel: 
$0.01<S<0.1$ cts~s$^{-1}$.
Middle panel: $0.1<S<1$ cts~s$^{-1}$. Bottom panel: $1<S<10$ cts~s$^{-1}$.
This example is included in the demo-version of the code.}
\label{label3}
\end{figure}

The kick velocity is not a crucial ingredient of our model, as we are dealing with young sources. 
It is important just that there is some wide distribution with a typical velocity about few hundreds km s$^{-1}$.
We choose for our code the distribution suggested by Arzoumanian et al. (2002). Tests showed that any other reasonable
distribution (e.g. Hobbs et al. 2005) discussed in the literature will work similarly well.

The mass spectrum (gravitational masses)
of NSs can be specified by the user.  As a default we use the spectrum
described in (Posselt et al.  2008).
It is based on calculations by Heger et al.  (2005) and on the HIPPARCOS
data on massive stars inside 500 pc from the Sun.
Note, that this is a mass spectrum for close-by young NSs, i.e. it can be
slightly different from the global NS mass spectrum in the Galaxy, as the
vicinity ($\sim 1$ kpc) of the Sun is strongly influenced by the presence of
the Gould Belt. We expect that the mass spectrum of near-by young NSs is
enriched with relatively low-mass objects ($\sim 1.1$-1.3 $M_\odot$).

Cooling curves are specified by the user.  
By default, we suggest to upload eight cooling curves for different masses
(also can be specified by the user), 
but the user may choose to upload a smaller
number of curves.  The code runs with a timestep of $10 000$ years.  If the
uploaded cooling curves have different timesteps, they are interpolated.  
As cooling curves are calculated for some particular EoS, then the user must
specify radii which correspond to each mass value used.  
Cooling curves can depend on different parameters in addition to the EoS,
for example, on the magnetic field (Pons, Geppert 2007, Pons et al. 2009).
As far as these parameters do not influence values used in the code, they
should not be specified, and are somehow ``hidden'' in cooling curves
properties. 

As our sources due to their young ages do not travel far from their birth places during the time of calculations, and
as we are interested in the region far from the Galactic centre,
then the model of the Galactic gravitational potential may not be extremely precise.
We use a simple three component (disc, bulge, halo) potential proposed by Miyamoto \& Nagai (1975) and modified by
Blaes \& Madau (1993). 

In our previous studies we used different approaches for the ISM distribution. 
Roughly, they can be divided into two groups: a simple smooth ``semianalytical distribution'' and  more detailed and ``patchy''
distributions based on many observations of individual sources in different directions.
It was shown by  Posselt et al. (2008) that for the Log~N~--~Log~S a simple distribution is good enough. 
The usage of detailed distributions make calculations much slower. So, for the web-tool we use the simple one.

We use the latest version of data for calculating the absorption in the ISM (Wilms et al. 2000). 
Our sources are soft (tens-hundreds of eV) and their emission is strongly absorbed. That is why absorption is one of the crucial ingredients
of the model.

All final results are given for the ROSAT PSPC, as we compare our calculations with the data obtained by this instrument. 
The Log~N~--~Log~S distribution, for example, is given in ROSAT counts per second.
Blackbody emission is assummed, and all influences of high magnetic fields
an/or atmospheres are neglected.

In Figure 1 we show the Log~N~--~Log~S for a set of cooling curves from Blaschke et al. (2004), their Fig. 21,
as it is calculated by the web-tool (the demo for this set of curves is available with the tool).
Dots represent 12 observed sources (see above). 
Error bars are poissonian. The dimmest source is the ``second Geminga''. At such low
fluxes, clearly, a significant fraction of ICoNSs is not identified,  
a good level of identification can be expected down to $\sim 0.1$
cts s$^{-1}$. 

Obviously, there are uncertainties in the model used. However, we consider that if a
Log~N~--~Log~S for some set of cooling
curves is systematically below or above the observational points outside the error bars, then such  a model
of thermal evolution is incorrect. Significant over- or under- prediction for bright sources ($>0.3$ cts s$^{-1}$) is also
a sign of serious problems with the cooling model.

In Figures 2-3 we demonstrate other types of data calculated by the population synthesis code.
This includes age and distance distributions for three intervals of ROSAT count rates
$S$
($0.01<S<0.1$; $0.1<S<1$; 
$1<S<10$), see detailed description in Posselt et al. (2008). All data files can be downloaded by the user. 
In addition, the user can download a map of the expected source distribution in galactic coordinates. However, as
we use a simple version of the three-dimensional ISM distribution in the web-tool, the map can be used just as a rough
illustration.

\section{User's manual}

The tool has clear web-forms, and it is very easy to operate it. 
From the welcome page of the web-tool (http://www.astro.uni-jena.de/Net-PSICoNS), the user can go to a demo-run, or 
to the main form, see Figure 4. 

Several demos are available for different sets of cooling curves.
The first demo-version is based on eight cooling curves calculated by Blaschke et al. (2004) (see
Figures
1-3).
The user can have a look at all output data: figures and files.
In additions, we present several demos based on cooling curves calculated by  Aguilera et al. (2008a, 2008b), Pons et al. (2009)
in the framework of additional heating due to magnetic field decay, and used
by Popov et al. (2010) for
population synthesis studies.

\begin{figure*}
\includegraphics[width=\textwidth]{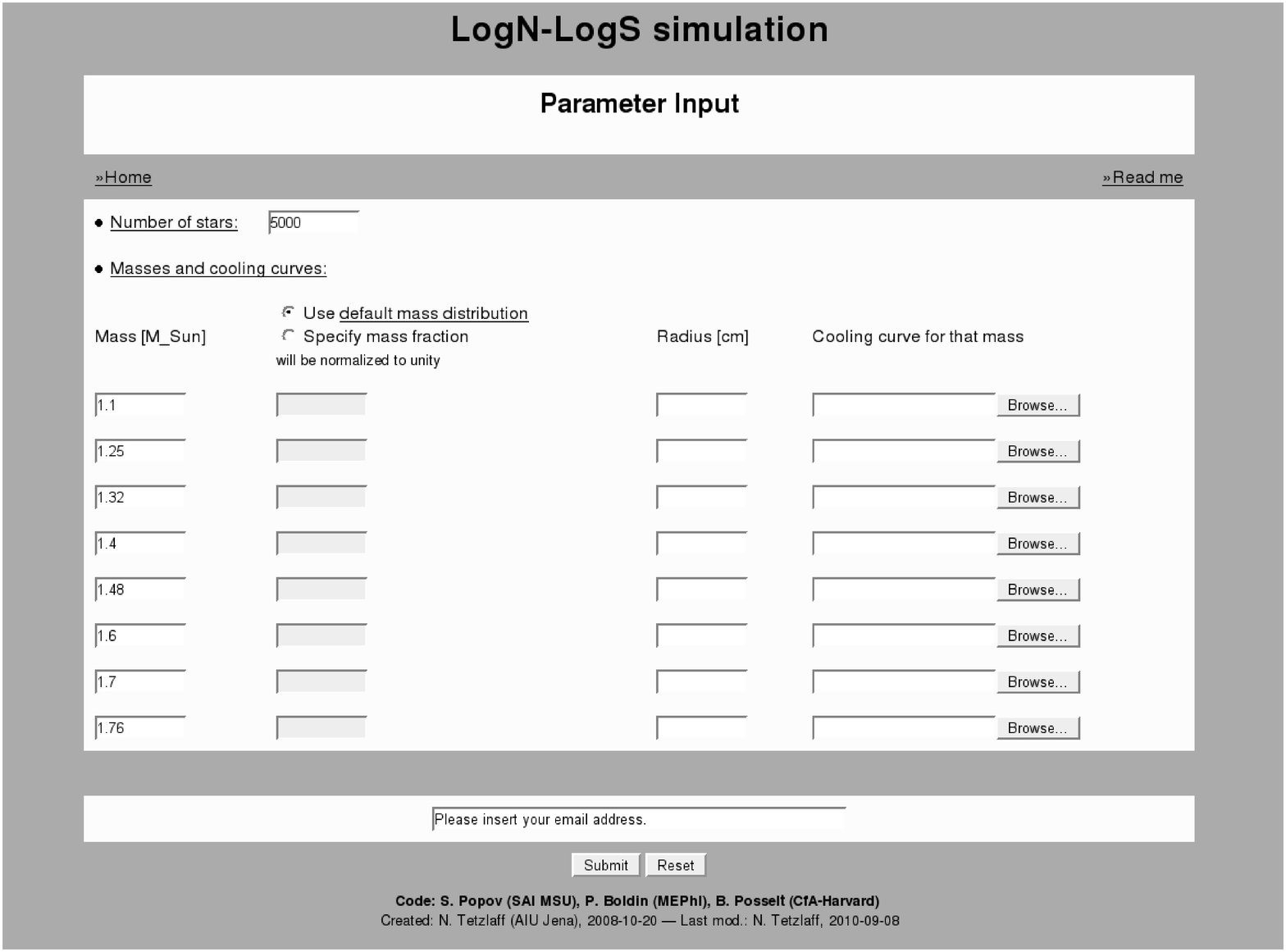}
\caption{The main form of the web-tool.}
\label{label4}
\end{figure*}

In the main form, the user has to specify the input data. For the number of stars in a run we suggest to use $\sim5000$. 
This is typically enough
to avoid all statistical fluctuations, unless the cooling track is very long. We suggest to
cut cooling curves
when the hottest NS reaches a temperature  of $\sim 300 000$~-~$100 000$~K. We recommend to stop at $100 000$K, 
unless the evolutionary track
becomes too long. Colder objects could not be detected by ROSAT
even at small distances. We do not recommend to use very long tracks. Already five million years is long enough for the
selected region with a radius of 3 kpc. Tracks with such a duration and longer tracks will provide only a lower limit
for Log~N~--~Log~S, also the calculations can take a rather long time. A cut at $\sim200 000$~-~$300 000$~K is a better option, if
NSs stay hotter than $100 000$~K for more than 5 Myrs.

Furthermore, the user has to define masses for cooling curves (up to 8) and specify the mass spectrum (if  not specified, then
it will be generated automatically from the default one). The final mass spectrum used in the actual calculations will be shown
to the user with the other output data.
Then, the user specifies radii for the chosen masses. And, finally, files with cooling curves have to be uploaded.

\begin{figure}
\includegraphics[width=80mm]{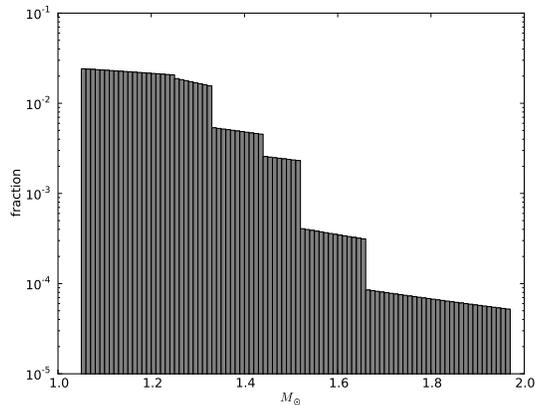}
\caption{The mass spectrum used if a user does not define a mass fraction by himself.
 For calculations the spectrum is re-bined for the given number of masses
(up to 8).}
\label{label5}
\end{figure}

Masses are always gravitational. Radii are circumferential. 
Calculations are made with a time step  of $10 000$ years. However, cooling curves
can have any (including non-uniform) timestep. They will  be automatically interpolated.
The minimum mass used in the mass spectrum is $1.05$ solar masses, the maximum mass is $1.96$ solar masses.
For the user-defined mass values the mass spectrum used in a run is re-calculated using our underlying spectrum (Fig. 5).
The general procedure to obtain  the NS mass distribution was  described in Popov et al. (2005, 2006a). 
Here we slightly adjusted it to allow re-binning. 
We use now an underlying spectrum with an $0.01$ solar mass step. There are six wide bins, which reflect the six spectral classes used 
(B2, B1, B0, O9, O8, O7) for NS progenitors.  Inside each wide bin we used the Salpeter mass function to re-distribute small 
($0.01$ solar mass)
bins. The mass spectrum used to calculate bin fractions for user-defined masses is shown in Figure 5. 

Our previous experience tells us that 8 cooling curves for different masses
is enough to cover the whole mass spectrum.
The user may also make calculations with less than 8 cooling curves.

The main population synthesis code is written in Fortran. 
All scripts are written in Python or Bash. 
The original Fortran code is not available for the user, but interested people can contact us on this topic.



\section{Conclusions}

We presented the web-tool \textsc{Net-PSICoNS} to calculate the Log~N~--~Log~S distribution for close-by ICoNSs using user-defined cooling curves.
The tool can be used mainly as a test for cooling curves providing an opportunity to make  a quick comparison 
of theoretical predictions with the observational data. 

Users comments about the tool are welcomed.

\section*{Acknowledgments}

We thank Dr. Bettina Posselt who was a co-author in several papers on the population synthesis of
isolated neutron stars. We thank Profs. David Blaschke and Jose Pons and their collaborators for providing cooling curves
for the default calculations. PB and SP thank the Observatory of Jena for hospitality.
PB and SP are supported by RFBR (09-02-00032, 10-02-00599) and Federal programme for
scientific and educational personnel (02.740.11.0575).
NT acknowledges financial support from Carl-Zeiss-Stiftung and DFG in the SFB/ TR-7 Gravitational
Wave Astronomy.



\begin{thebibliography}{}
\bibitem{}
Aguilera, D. N., Pons, J. A., Miralles, J. A.: 2008a, ApJ 673, L167
\bibitem{}
Aguilera, D. N., Pons, J. A., Miralles, J. A.: 2008b, A\&A 486, 255
\bibitem{}
Arzoumanian, Z., Chernoff, D.F., Cordes, J.M.: 2002, ApJ 568, 289
\bibitem{}
Blaes, O., Madau, P.: 1993, ApJ 403, 690 
\bibitem{}
Blaschke, D., Grigorian, H., Voskresensky, D.: 2004, A\&A 424, 979
\bibitem{}
Haberl, F.: 2007, Ap\&SS 308, 181 
\bibitem{}
Haensel, P., Potekhin, A.Y., Yakovlev, D.G.: 2007,
``Neutron Stars 1 : Equation of State and Structure'', Astrophysics and space science library, Vol. 326.  New York: Springer
\bibitem{}
Heger, A., Woosley, S.E., Spruit, H.C.: 2005, ApJ 626, 350
\bibitem{}
Hobbs, G., Lorimer, D. R., Lyne, A. G., Kramer, M.: 2005, MNRAS 360, 974
\bibitem{}
Miyamo, M., Nagai, R.: 1975, PASJ 27, 533
\bibitem{}
Pizzochero, P.M.:2010, arXiv:1001.1272
\bibitem{}
Pons, J., Geppert, U.: 2007, A\&A 470, 303
\bibitem{}
Pons, J.A., Miralles J. A., Geppert U.: 2009, A\&A, 496, 207
\bibitem{}
Popov, S.B., Colpi, M., Prokhorov, M.E.,  Treves, A., Turolla, R.: 2003, A\&A 406, 111
\bibitem{}
Popov, S.B., Turolla, R., Colpi, M., Prokhorov, M.E., Treves, A.: 2005, Ap\&SS 299, 117
\bibitem{}
Popov, S.B., Grigorian, H., Blaschke, D.: 2006b, Phys. Rev. C 74, 025803
\bibitem{}
Popov, S.B., Pons, J.A., Miralles, J.A., Boldin, P.A., Posselt, B.: 2010, MNRAS 401, 2675
\bibitem{}
Popov, S.B., Grigorian, H., Turolla, R., Blaschke, D.: 2006a, A\&A 448, 327
\bibitem{}
Popov, S.B., Prokhorov, M.E.: 2007, Physics Uspekhi 50, 1123
\bibitem{}
P\"oppel, W.: 1997, Fund. Cosm. Phys. 18, 1
\bibitem{}
Posselt, B, Popov, S.B., Haberl, F., Truemper, J., Turolla, R., Neuhauser, R.: 2008, A\&A 482, 617
[see Erratum in: A\&A 512, id. C2]
\bibitem{}
Prakash, M., Lattimer, J.M., Pons, J.A., Steiner, A.W., Reddy, S.:
2001, in: ``Physics of neutron star interiors'',
eds. Blaschke D., Glendenning N.K., Sedrakian A., p.364 [ArXiv:
astro-ph/0012136]
\bibitem{} 
Ridley, J.P., Lorimer, D.R.: 2010, MNRAS 404, 1081
\bibitem{}
Schmitt, A.: 2010, Lecture notes in Physics (in press), [arXiv:1001.3294]
\bibitem{}
Wilms, J., Allen, A., McCray, R.: 2000, ApJ, 542, 914
\bibitem{}
Zane, S. et al.: 2010, MNRAS (in press), [arXiv: 1009.0209]
\end{thebibliography}
\end{document}